\documentclass{kapproc}
\usepackage{psfig}

\begin{document}

\articletitle[]{How to determine the star formation histories in spiral
disks}
\chaptitlerunninghead{Determining the star formation histories in disks}
\author{Mercedes Moll\'{a}}
\affil{Dpto. de F\'{\i}sica Te\'{o}rica C-XI, 
Universidad Aut\'{o}noma de Madrid, 28049 Madrid, Spain}
\email{mercedes.molla@uam.es}

\author{Eduardo Hardy\footnote{
The National Radio Astronomy Observatory is a facility of the National
Science Foundation operated under cooperative agreement by Associated
Universities, Inc.}}
\affil{NRAO, Casilla 36-D, Santiago, Chile}
\email{ehardy@nrao.edu}

\section{Introduction}

With the help of the multiphase chemical evolution model (Moll\'{a},
Hardy \& Beauchamps 1999) we have derived the evolutionary histories
of the galaxies NGC 4303, NGC 4321 y NGC 4535. With these histories
and an evolutionary synthesis model, we were able to reproduce their
observed radial distributions of the spectral indices Mg2 and Fe5270.

Chemical evolution models however may fit well the present day
observational characteristics of galaxies without discriminating among
very different star formation histories.  Furthermore, the spectral
indices Mg2 and Fe5270 exhibit the age-metallicity degeneracy
problem, implying that discriminating ages from metallicity with only
these indices in single stellar populations might be impossible. Our
objective here is to make sure that the implied evolutionary histories
represent well these galaxies.

\section{The multiphase model}

The multiphase scenario begins with a gaseous protogalaxy whose mass
is calculated from a rotation curve. The gas collapses onto the
equatorial plane, thus forming the disk, at a rate which depends on the
total mass: $
\frac{\tau_{0,gal}}{\tau_{0,MWG}}=[\frac{M_{MWG}}{M_{gal}}]^{1/2}$

The gas collapses more rapidly in the central regions than in the
outer disk, so we assume: $\tau_{coll}(\rm R)=\tau_{0}\exp{((\rm
R-R_{0})/\lambda_{D})}$.  Thus, we must select two parameters,
$\tau_{0,gal}$ and $\lambda_{D}$, for a particular galaxy model.

Stars form in the halo from the primordial gas following a Schmidt
law. In the disk, molecular clouds form from the diffuse gas while
stars form in a second step as a result of cloud-cloud
collisions. Stars also form via the interaction of massive stars
with molecular clouds. The former is a local process governed by an
efficiency constant valid for all galaxies. The same is valid for the
halo star formation efficiency.  Thus, for a given galaxy only the two
efficiencies, namely $\epsilon_{\mu}$ and $\epsilon_{H}$, must be
chosen in order to compute a model. They are larger for earlier
morphological types than for the later ones, being well represented by
a probability function depending on the Hubble Type T: $\epsilon_{\mu}
= e^{-T^{2}/15}$ and $\epsilon_{H} = e^{-T^{2}/5}$ (see Moll\'{a},
D\'{\i}az \& Ferrini 2002 for details).
 
Summarizing, to run a model for a galaxy, {\sl i.e} NGC 4303, we need
3 input parameters: 1) The collapse time scale $\tau_{0}$, determined
by the total mass obtained from V(R); 2) The Hubble type T, with which
we determine the set of efficiencies ($\epsilon_{\mu}, \epsilon_{H}$);
3) The scale length $\lambda_{D}$. 

\section{The uniqueness problem of models for NGC 4303}

The problem is that uncertainties exist in the selection of the above
parameters. For the galaxy NGC~ 4303, the optical and radio rotation
curves look different: the maximum rotation velocity is a factor 1.4
larger in the second case. This in turn implies that the collapse time
scale may be different than the one selected ($\sim 6$ Gyr) in the
first case.  It follows that the efficiencies will also have to change
in order to fit the present day observational data. Thus, alternative
models would seem possible.

We have varied these input parameters within a reasonable range
centered around those values selected in Moll\'{a} et al. (1999) for
NGC~4303. Models with all possible combinations of $\tau_{0}=$1, 4, 8,
12 and 16 Gyr, $\lambda_{D}$=1,4,8,12,and 16 kpc, and T from 1 to 10
with a variation step of 0.5, that is 20 possible values for the set
($\epsilon_{\mu}, \epsilon_{H}$), have been ran, thus doing a total of 500
models. (See details in Moll\'{a} \& Hardy 2002).

\section{The $\chi^{2}$ method}

In order to select the best models out of the above 500, we compare
the radial distributions for the oxygen, diffuse and molecular gas,
and star formation rate resulting form our models with the
observations.

This comparison is performed using the statistical indicator
$\chi^{2}$, which measures the goodness of fit via the differences
between the model results and the data, taking into account the
measured dispersion of data.We show in Table~\ref{sel} the
characteristics and parameters of models falling within the 97.5~\%
confidence region (i.e., a 90\% of the combined values).

We display in column (1) the number identifier for the model. The
characteristics collapse time scale, $\tau_{0}$, and the scale length,
$\lambda_{D}$, are in columns (2),and (3) respectively. Column (4)
contains the value of T.  The $\epsilon_{\mu}$ and $\epsilon_{H}$
efficiencies are in columns (5) and (6).
\begin{table}
\caption{Parameters of the Selected Models (P $>$ 90\%)}
\begin{tabular*}{\hsize}{@{\extracolsep{\fill}}rccccccccccc}
\sphline
 Model & $\tau$ & $\lambda_{d}$ & T & $\epsilon_{\mu}$ &$\epsilon_{H}$
& P$_{\rm OH    }$        &P$_{\rm SFR  }$
& P$_{\rm H_{I} }$        &P$_{\rm H_{2}}$
&P$_{\rm Mg_{2}}$         &P$_{\rm Fe52}$ \\
Number& Gyr & kpc & & & & & & & & &  \\
\sphline
$\ast$ 129&  4. & 4. &   4.5& .259&.017&    1.00&  0.98&  1.00&  1.00 & 1.00&  1.00\cr
$\ast$ 228&  8. & 4. &   4.0& .344&.041&    1.00&  1.00&  1.00&  0.98 & 1.00&  1.00\cr
$\ast$ 229&  8. & 4. &   4.5& .259&.017&    1.00&  1.00&  1.00&  1.00 & 1.00&  1.00\cr
       248&  8. & 8. &   4.0& .344&.041&    0.99&  0.99&  1.00&  0.99 & 0.97&  1.00\cr
$\ast$ 249&  8. & 8. &   4.5& .259&.017&    1.00&  0.99&  1.00&  1.00 & 0.98&  1.00\cr
$\ast$ 329& 12. & 4. &   4.5& .259&.017&    1.00&  0.98&  1.00&  1.00 & 1.00&  1.00\cr
       348& 12. & 8. &   4.0& .344&.041&    1.00&  1.00&  1.00&  0.99 & 0.95&  1.00\cr
$\ast$ 349& 12. & 8. &   4.5& .259&.017&    1.00&  1.00&  1.00&  1.00 & 0.98&  1.00\cr
       368& 12. &12. &   4.0& .344&.041&    0.99&  0.99&  1.00&  0.99 & 0.91&  1.00\cr
       369& 12. &12. &   4.5& .259&.017&    1.00&  0.99&  1.00&  1.00 & 0.88&  1.00\cr
       388& 12. &16. &   4.0& .344&.041&    0.99&  0.98&  1.00&  0.99 & 0.88&  1.00\cr
       389& 12. &16. &   4.5& .259&.017&    1.00&  0.98&  1.00&  0.99 & 0.90&  1.00\cr
       448& 16. & 8. &   4.0& .344&.041&    1.00&  0.99&  1.00&  0.99 & 0.94&  1.00\cr
       449& 16. & 8. &   4.5& .259&.017&    1.00&  1.00&  1.00&  1.00 & 0.83&  0.90\cr
       450& 16. & 8. &   5.0& .189&.007&    1.00&  1.00&  0.99&  1.00 & 0.83&  0.90\cr
       468& 16. &12. &   4.0& .344&.041&    1.00&  0.99&  1.00&  0.99 & 0.90&  1.00\cr
       469& 16. &12. &   4.5& .259&.017&    1.00&  1.00&  1.00&  1.00 & 0.93&  0.99\cr
       488& 16. &16. &   4.0& .344&.041&    0.99&  0.99&  1.00&  0.99 & 0.87&  1.00\cr
       489& 16. &16. &   4.5& .259&.017&    1.00&  0.99&  1.00&  1.00 & 0.92&  0.99\cr
\sphline
\end{tabular*}
\label{sel}
\end{table}

The probabilities of these
distributions to be in a region around the minimum value of
$\chi^{2}$, for each one of our observational constraints, are listed
in columns (7) to (10). By selecting the best models as those falling
within the 90\% confidence region we reduce the number of possible
models to 19.  All other models have probabilities smaller than this,
at least in one of the 4 observational constraints.

\section{The evolutionary synthesis models}

In order to derive integrated stellar abundances, we calculate the
integrated mass of all stars created in a given time interval, and the
mean abundance reached at that epoch by the gas out of which they
form. We consider the stellar populations residing at every
galactocentric region as the superposition of a set of single stellar
populations or {\sl generations} each one defined by its age and its
metallicity.  Spectral index features are calculated with the same
method described in Moll\'{a} \& Garc\'{\i}a-Vargas (2000) by using
the Padova group isochrones. The fitting functions from Worthey (1994)
and from Idiart et al (1997), for assigning the index Fe52 and
Mg$_{2}$, respectively, to each star are used.  We obtain this way the
radial distributions of these spectral indices for the 19 models of
Table~\ref{sel}.

As before, we compute the $\chi^{2}$, by comparing these two radial
distributions with the observations (Beauchamp \& Hardy 1997;
Moll\'{a} et al. 1999). The resulting probabilities are given in
Columns (11) and (12) of Table~\ref{sel}. Only 6 of the 19 models,
marked with an $\ast$ in this same table, fit these new constraints
with probabilities larger than 97.5~\%.

\section{Conclusions}

We have computed 500 models for the galaxy NGC 4303 with different
input parameters. By using the usual goodness-of-fit chi-square
parameter we select those reproducing the present-day data within a
confidence level of 90\%, and we constraint the possible number to 19.
However, these models do not reproduce equally well the spectral index
radial distributions. Out of these 19 models, only 6 are {\it also}
able to reproduce the radial distributions of spectral indices.

It seems that the uniqueness problem associated to chemical evolution
models is not strong: possible models reduce to 5\% of the initial
ones using only the present day data as constraints. When we also use
the spectrophotometric indices, we limit even more the possible
evolutionary histories, by reducing to a  1\% the possible models
out of the initial 500.
 
We conclude that this technique combining chemical evolution with
evolutionary synthesis is a very powerful tool to discriminate among
evolutionary scenarios in galaxies with continuous star formation. We
propose more observational campaigns in order to obtain the present as
well as other spectral indices in spiral and irregular galaxies.


\begin{chapthebibliography}{}

\bibitem[]{}Beauchamp, D., \& Hardy, E. 1997, AJ, 113, 1666 (Paper I)
\bibitem[]{}Ferrini, F., Palla F., \& Penco, U. 1990, A\&A, 213, 3
\bibitem[]{}Idiart, T. P., \& de Freitas-Pacheco, J. A. 1995, AJ, 109, 2218
\bibitem[]{}Moll\'{a}, M., Hardy, E., \& Beauchamp, D.  1999, ApJ, 513, 695 
\bibitem[]{}Moll\'{a}, M. \& Hardy, E. 2002, AJ, 123, 3055
\bibitem[]{}Moll\'{a}, M. \& Garc\'{\i}a-Vargas, M.L. 2000, A\&A, 359, 18
\bibitem[]{}Moll\'{a}, M. D\'{\i}az, A.I. \& Ferrini, F., 2002, ApJ, submitted
\bibitem[]{}Worthey, G. 1994, ApJS, 95, 107
\end{chapthebibliography}
\end{document}